\documentclass{iaus}
\usepackage{graphicx}

\title[First Stars] %% give here a short title %%
{Formation of the First Stars}

\author[V. Bromm]   %% give here a short author list %%
{Volker Bromm$^1$}

\affiliation{$^1$Department of Astronomy, University of Texas,
Austin, TX 78712, USA \break email: vbromm@astro.as.utexas.edu\\[\affilskip]}

\pubyear{2005}
\volume{228}  %% insert here the IAU Symposium No.
\pagerange{1--10}
\date{?? and in revised form ??}
\jname{From Lithium to Uranium: Elemental Tracers of Early Cosmic Evolution}
\editors{V. Hill, P. Fran\c{c}ois \& F. Primas, eds.}
\begin{document}

\maketitle

\begin{abstract}
How and when did the first generation of stars form at the
end of the cosmic dark ages? Quite generically,
within variants of the cold dark matter model of cosmological
structure formation, the first sources of light are expected
to form in $\sim 10^{6} M_{\odot}$ dark matter potential wells
at redshifts $z\geq 20$.
I discuss the physical processes that govern the formation of
the first stars. These so-called Population~III stars are predicted
to be predominantly very massive, and to
have contributed significantly to the early reionization of the 
intergalactic medium. 
Such an early reionization epoch is inferred from the recent
measurement of the Thomson optical depth by the {\it WMAP} satellite.
I address the importance of heavy elements in bringing
about the transition from an early star formation mode dominated by
massive stars, to the familiar mode dominated by low mass stars, at
later times, and present possible observational probes.
This transition could have been gradual, giving rise to
an intermediate-mass population of still virtually metal-free stars
(``Population II.5''). These stars could have given rise to the
peculiar class of black-hole forming supernovae inferred from
the abundance pattern of extremely iron-poor stars.
\keywords{cosmology: theory, stars: formation, supernovae: general}
\end{abstract}

\firstsection % if your document starts with a section,
\section{Introduction}

One of the grand challenges in modern cosmology is posed by the
question:
How did the first stars in the universe form, what were their
physical properties, and what was their impact on cosmic history
(e.g., Bromm \& Larson 2004)?
The first stars, formed at the end of the cosmic dark ages,
ionized (e.g., Wyithe \& Loeb 2003; Cen 2003) and metal-enriched
(e.g., Furlanetto \& Loeb 2003) the intergalactic medium (IGM) and consequently had
important effects on subsequent galaxy formation (e.g., Barkana \& Loeb 2001) and on
the large-scale polarization anisotropies of the cosmic microwave
background (Kaplinghat et al. 2002).  {\it When did the cosmic dark ages end?}  In
the context of popular cold dark matter (CDM) models of hierarchical
structure formation, the first stars are predicted to have formed in
dark matter halos of mass $\sim 10^{6}M_{\odot}$ that collapsed at
redshifts $z\simeq 20-30$ (e.g., Barkana \& Loeb 2001; Yoshida et al. 2003). 

Results from recent numerical simulations of the collapse and
fragmentation of primordial clouds suggest that the first stars were
predominantly very massive, with typical masses $M_{\ast}\ge 100
M_{\odot}$ (Bromm, Coppi, \& Larson 1999, 2002; Nakamura \& Umemura 2001;
Abel, Bryan, \& Norman 2002).  Despite the
progress already made, many important questions remain unanswered.
An example for an open question is: {\it How
does the primordial initial mass function (IMF) look like?}  Having
constrained the characteristic mass scale, still leaves undetermined
the overall range of stellar masses and the power-law slope which is
likely to be a function of mass. In addition, it is presently unknown
whether binaries or, more generally, clusters of zero-metallicity
stars, can form.  
{\it What is the nature of the feedback that the first stars exert on
their surroundings?}  The first stars are expected to produce copious
amounts of UV photons and to possibly explode as energetic
hypernovae. These negative feedback effects could 
suppress star formation in neighboring high-density clumps.

Predicting the properties of the first sources of light, in particular
their expected luminosities and spectral energy distributions, is important
for the design of upcoming instruments, such as the {\it James Webb Space
Telescope} (JWST) \footnote{See http:// ngst.gsfc.nasa.gov.}, or the
next generation of large ($>10$m) ground-based telescopes.  The hope
is that over the upcoming decade, it will become possible to confront
current theoretical predictions about the properties of the first
stars with direct observational data.

\section{Population~III star formation}

The metal-rich chemistry, magnetohydrodynamics, and radiative
transfer involved in present-day star formation is complex, and we
still lack a comprehensive theoretical framework that predicts the IMF
from first principles (see Larson~2003 for a recent review).
Star formation in the high redshift universe,
on the other hand, poses a theoretically more tractable problem due to
a number of simplifying features, such as: (i) the initial absence of
heavy metals and therefore of dust; and (ii) the absence of
dynamically-significant magnetic fields, in the pristine gas left over
from the big bang. The cooling of the primordial gas does then only
depend on hydrogen in its atomic and molecular form.  Whereas 
the initial state of the star forming
cloud is poorly constrained in the
present-day interstellar medium, 
the corresponding initial conditions for
primordial star formation are simple, given by the popular
$\Lambda$CDM model of cosmological structure formation. 

\begin{figure*}[t]
  \includegraphics[height=.3\textheight]{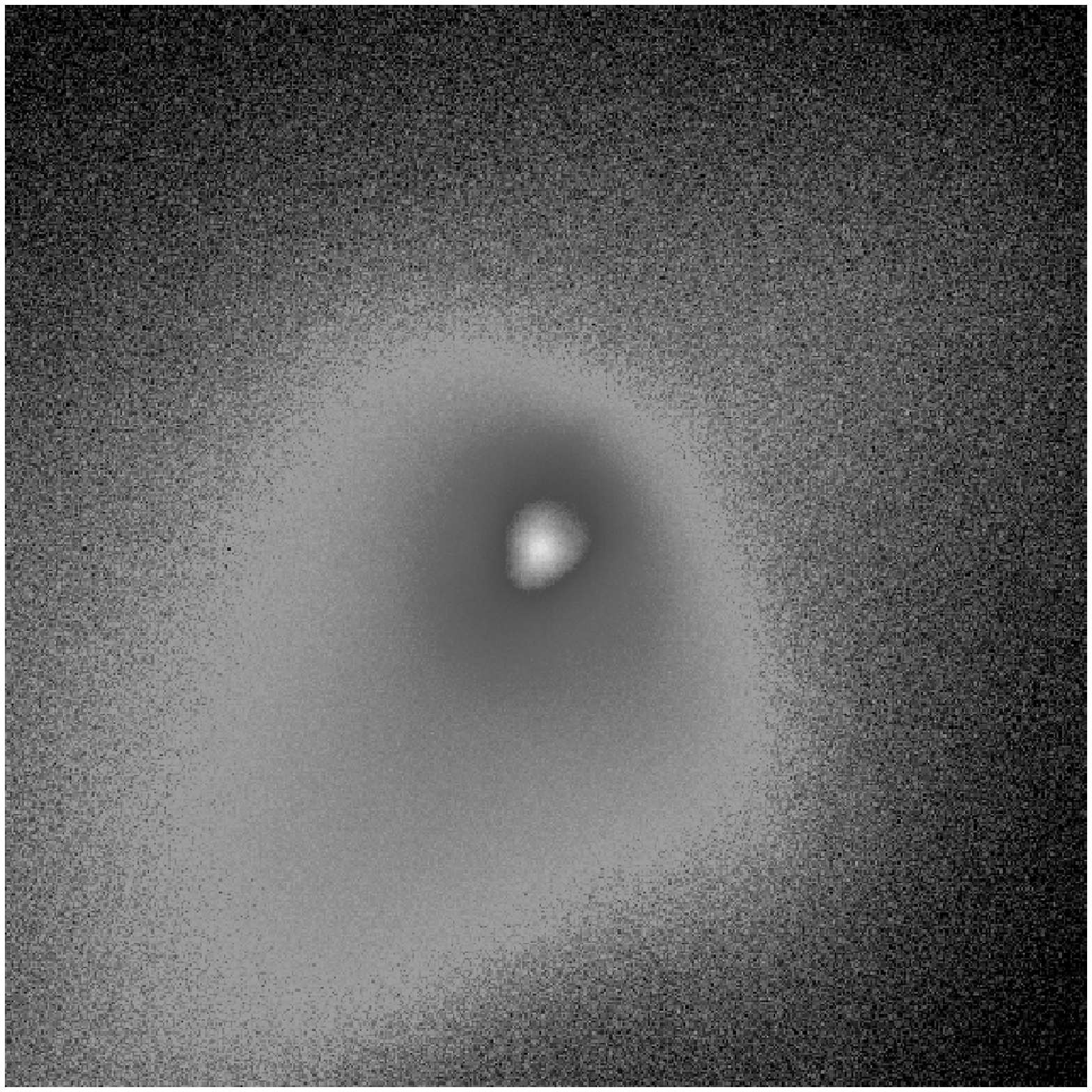}
  \includegraphics[height=.3\textheight]{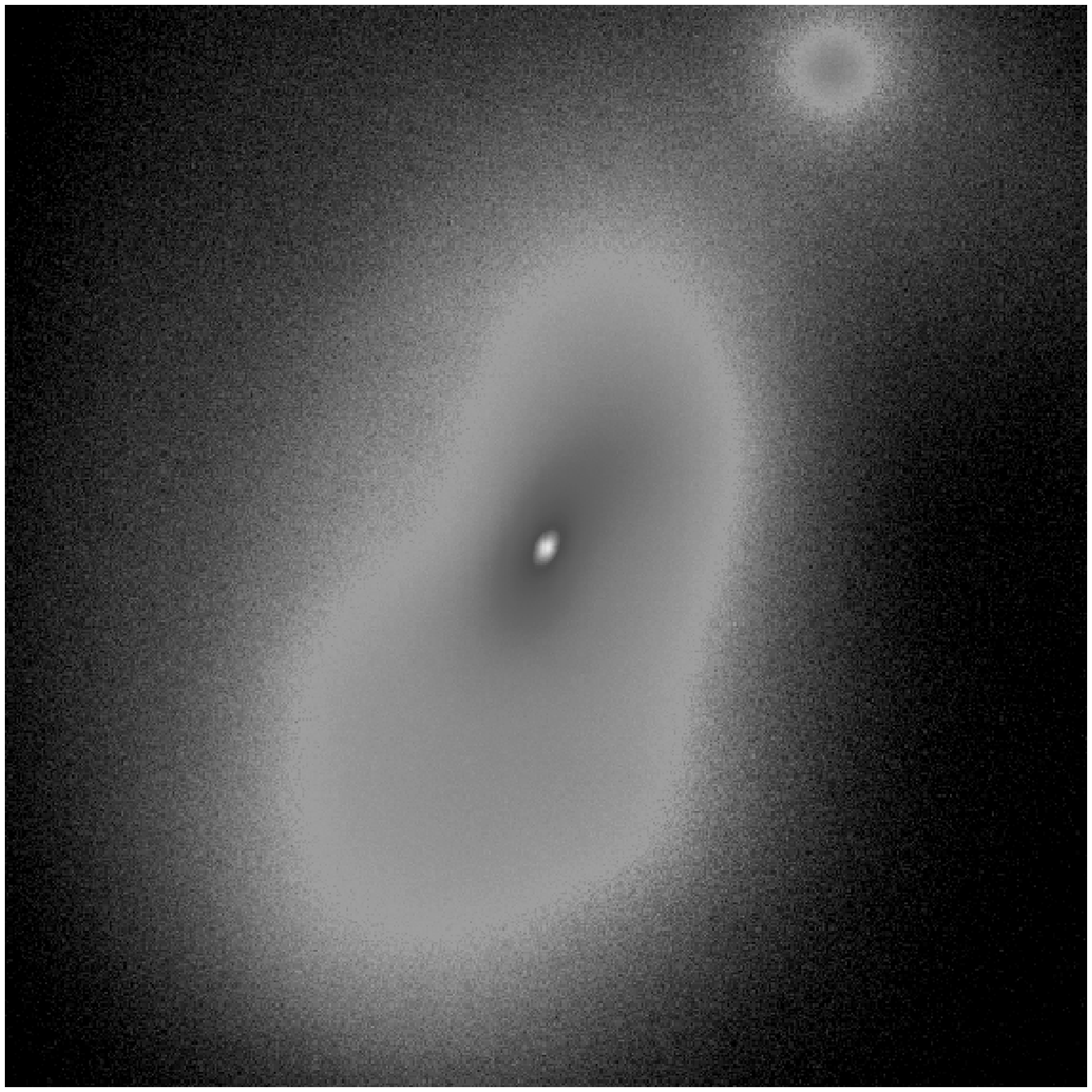}
\caption{Collapse and fragmentation of a primordial cloud.  Shown is
the projected gas density at a redshift $z\simeq 21.5$, briefly after
gravitational runaway collapse has commenced in the center of the
cloud.  {\it Left:} The coarse-grained morphology in a box with linear
physical size of 23.5~pc.  
{\it Right:} The fine-grain morphology in a box with linear physical
size of 0.5~pc.
The central density peak, vigorously gaining mass by accretion, is
accompanied by a secondary clump.
(Adapted from Bromm \& Loeb 2004.)}
\end{figure*}

\begin{figure*}[b]
  \includegraphics[height=.3\textheight]{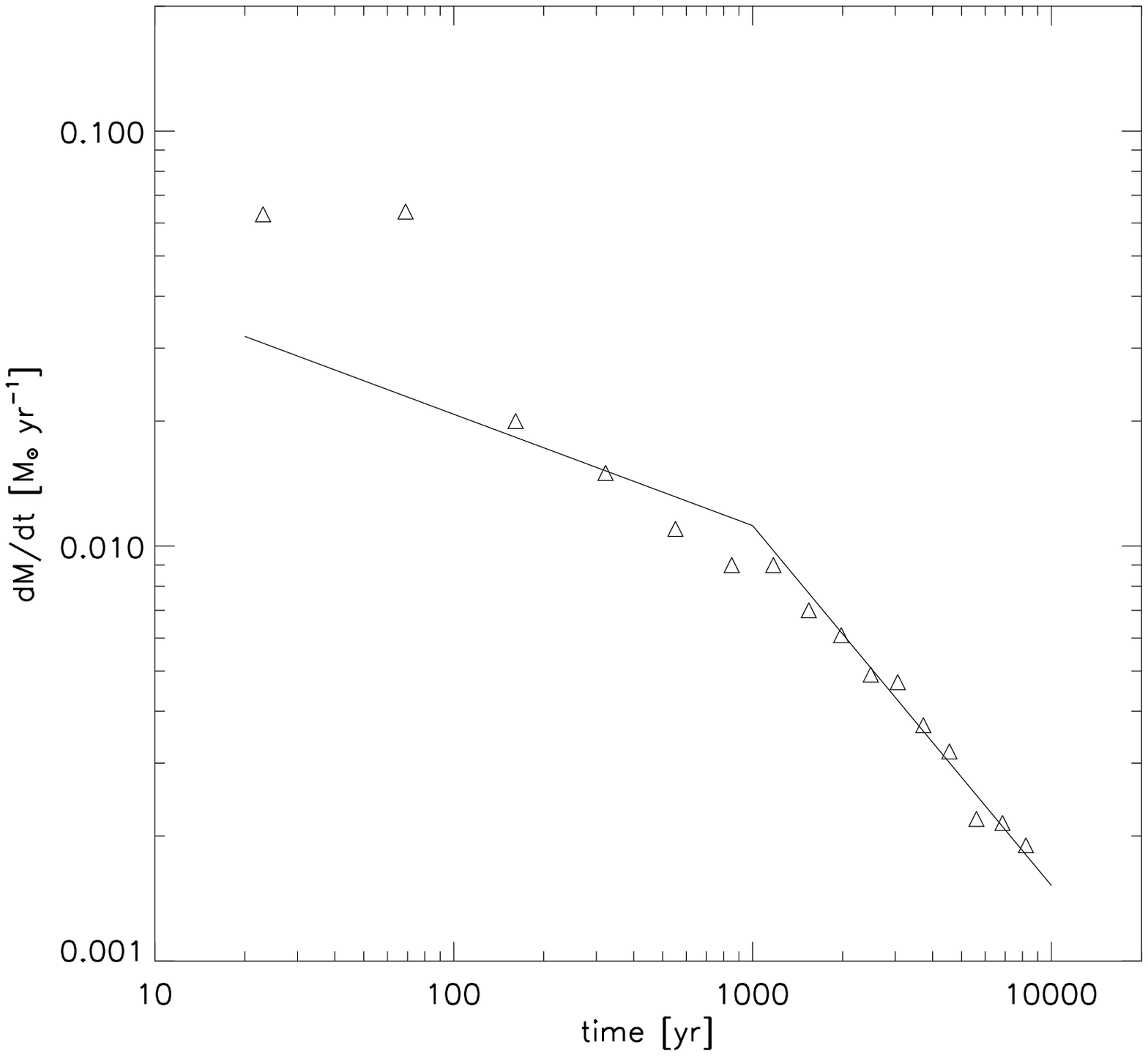}
  \includegraphics[height=.3\textheight]{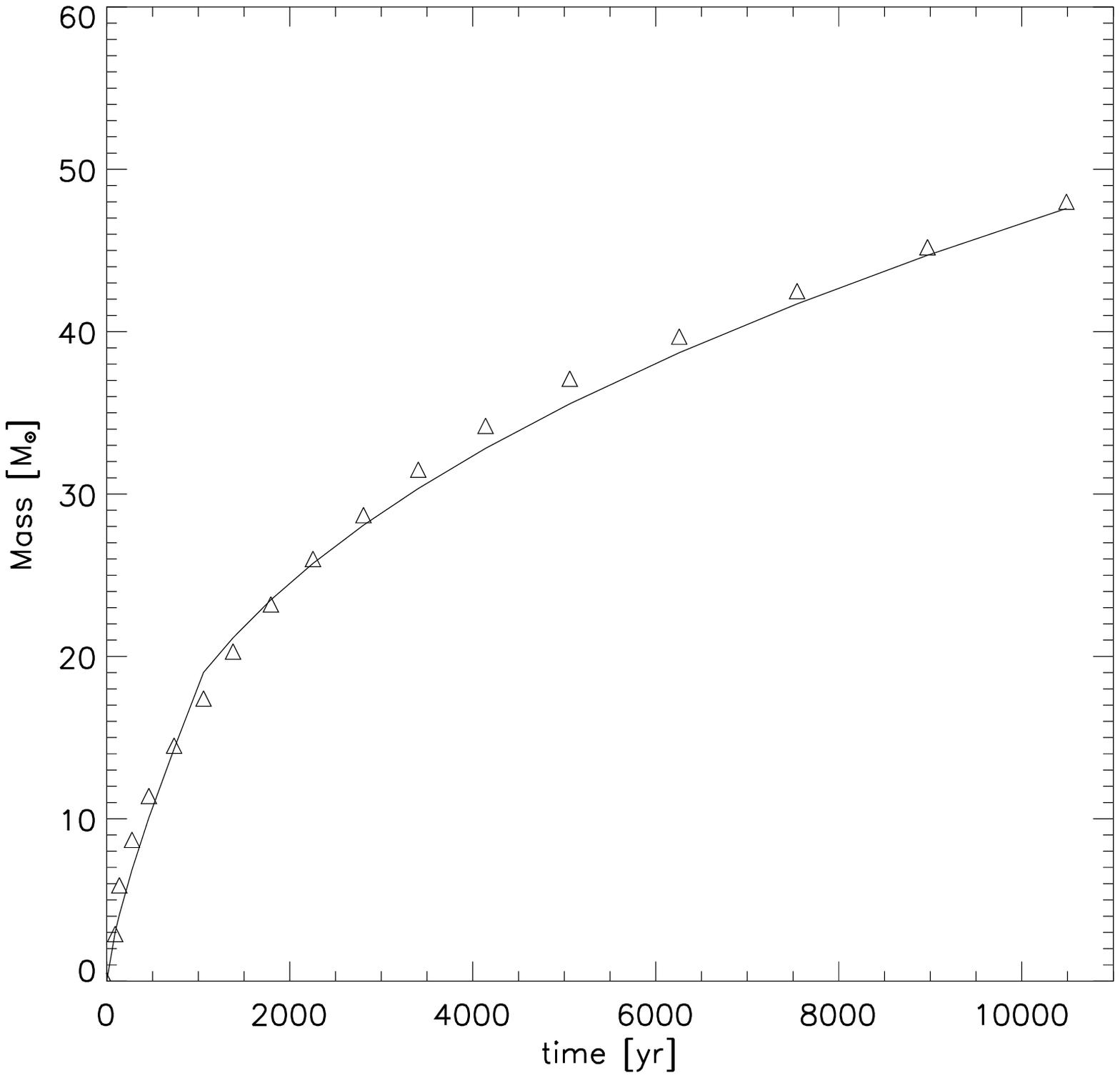}
\caption{Accretion onto a primordial protostar.  The morphology of
this accretion flow is shown in Fig.~1.  {\it Left:} Accretion rate
(in $M_{\odot}$~yr$^{-1}$) vs. time (in yr) since molecular core
formation.  {\it Right:} Mass of the central core (in $M_{\odot}$)
vs. time.  {\it Solid line:} Accretion history approximated as:
$M_{\ast}\propto t^{0.75}$ at $t<10^{3}$~yr, and
$M_{\ast}\propto t^{0.4}$ afterwards.
Using this analytical approximation, we
extrapolate that the protostellar mass has grown to $\sim 120
M_{\odot}$ after $\sim 10^{5}$~yr, and to $\sim 500 M_{\odot}$ after
$\sim 3\times 10^{6}$~yr, the total lifetime of a very massive star.
(Adapted from Bromm \& Loeb 2004.)}
\end{figure*}

{\it How did the first stars form?} A complete answer to this question
would entail a theoretical prediction for the Population~III IMF,
which is rather challenging. Let us start by addressing the simpler
problem of estimating the characteristic mass scale of the first
stars. This mass scale is observed to be $\sim 1
M_{\odot}$ in the present-day universe.  To investigate the collapse
and fragmentation of primordial gas, we have carried out numerical
simulations, using the smoothed particle hydrodynamics (SPH)
method. We have included the chemistry and cooling physics relevant
for the evolution of metal-free gas (see Bromm et al. 2002 for
details). Improving on earlier work (Bromm et al. 1999, 2002) by
initializing our simulation according to the $\Lambda$CDM model, we
focus here on an isolated overdense region that corresponds to a
3$\sigma-$peak: a halo containing a total mass of $10^{6}M_{\odot}$,
and collapsing at a redshift $z_{\rm vir}\simeq 20$ (Bromm \& Loeb 2004).
In Figure~1 ({\it left panel}), we show the gas density within the
central $\sim 25$~pc, briefly after the first high-density clump has
formed as a result of gravitational runaway collapse.

{\it How massive were the first stars?} Star formation typically
proceeds from the `inside-out', through the accretion of gas onto a
central hydrostatic core.  Whereas the initial mass of the hydrostatic
core is very similar for primordial and present-day star formation
(Omukai \& Nishi 1998), the accretion process -- ultimately responsible for
setting the final stellar mass, is expected to be rather different. On
dimensional grounds, the accretion rate is simply related to the sound
speed cubed over Newton's constant (or equivalently given by the ratio
of the Jeans mass and the free-fall time): $\dot{M}_{\rm acc}\sim
c_s^3/G \propto T^{3/2}$. A simple comparison of the temperatures in
present-day star forming regions ($T\sim 10$~K) with those in
primordial ones ($T\sim 200-300$~K) already indicates a difference in
the accretion rate of more than two orders of magnitude.

Our high-resolution simulation enables us to study the three-dimensional
accretion flow around the protostar (see also Omukai \& Palla 2001, 2003;
Ripamonti et al. 2002; Tan \& McKee 2003).  We allow the gas to reach
densities of $10^{12}$ cm$^{-3}$ before being incorporated into a
central sink particle. At these high densities, three-body reactions
(Palla, Salpeter, \& Stahler 1983) have converted the gas into a fully molecular form.  In
Figure~2, we show how the molecular core grows in mass over the first
$\sim 10^{4}$~yr after its formation. The accretion rate ({\it left
panel}) is initially very high, $\dot{M}_{\rm acc}\sim 0.1
M_{\odot}$~yr$^{-1}$, and subsequently declines with time.
The mass of the
molecular core ({\it right panel}), taken as an estimator of the
proto-stellar mass, grows approximately as: $M_{\ast}\sim \int
\dot{M}_{\rm acc}{\rm d}t \propto t^{0.75}$ at $t<10^{3}$~yr, and
$M_{\ast} \propto t^{0.4}$ afterwards.
A rough upper limit for
the final mass of the star is then: $M_{\ast}(t=3\times 10^{6}{\rm
yr})\sim 500 M_{\odot}$. In deriving this upper bound, we have
conservatively assumed that accretion cannot go on for longer than the
total lifetime of a very massive star (VMS).

{\it Can a Population~III star ever reach this asymptotic mass limit?}
The answer to this question is not yet known with any certainty, and
it depends on whether the accretion from a dust-free envelope is
eventually terminated by feedback from the star (e.g., Omukai \& Palla 2001,
2003; Omukai \& Inutsuka 2002; Ripamonti et al. 2002; Tan \& McKee 2003). The standard mechanism by which
accretion may be terminated in metal-rich gas, namely radiation
pressure on dust grains (e.g., Wolfire \& Cassinelli 1987), is evidently not effective for
gas with a primordial composition. Recently, it has been speculated
that accretion could instead be turned off through the formation of an
H~II region (Omukai \& Inutsuka 2002), or through the radiation pressure exerted
by trapped Ly$\alpha$ photons (Tan \& McKee 2003). The termination of the
accretion process defines the current unsolved frontier in studies of
Population~III star formation.

\section{Second generation stars}

\begin{figure*}[t]
  \includegraphics[height=.4\textheight]{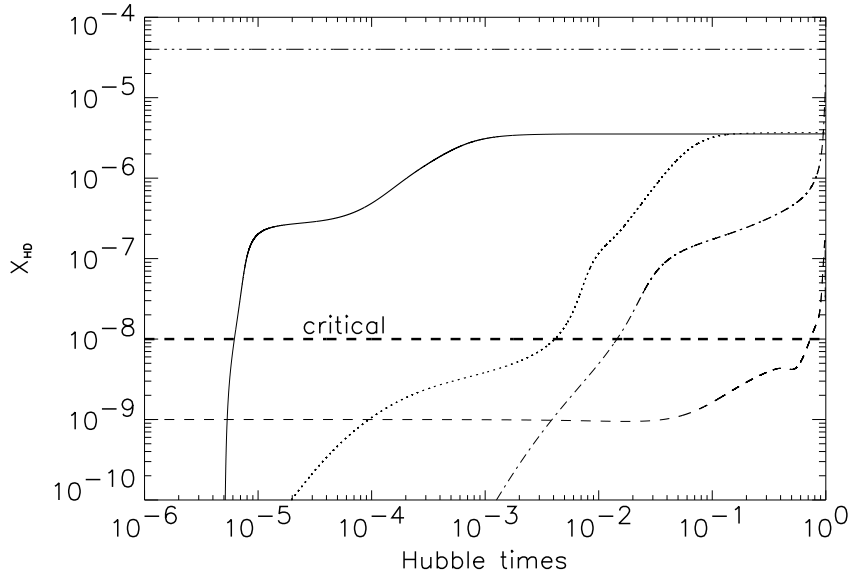}
\caption{Evolution of HD abundance in four distinct cases. 
{\it Solid line:} Gas compressed and heated by a SN explosion with
$u_{\rm sh}$= 100 km~s$^{-1}$.
{\it Dotted line:} Gas shocked in the build-up of a 3$\sigma$ fluctuation
dark matter halo collapsing at $z\simeq 15$.
{\it Dash-dotted line:} Gas collapsing inside a relic HII region,
which is left behind after the death of a very massive Pop~III star.
{\it Dashed line:} Gas collapsing inside a minihalo at $z\simeq 20$.
In this case, the gas does not experience a strong shock, and is never
ionized. Contrary to the other three cases, where the gas went through
a fully ionized phase, HD cooling is not important here. 
The critical HD abundance, shown by the bold dashed line, is defined such
that primordial gas is able to cool to the CMB temperature within the
fraction of a Hubble time. The CMB sets a minimum floor to the gas temperature,
because radiative cooling below this fllor is thermodynamically not 
possible. The HD abundance exceeds the critical value in a time which is
short compared to the Hubble time for all fully-ionized, strongly shocked
cases.
(Adapted from Johnson \& Bromm 2005.)}
\end{figure*}

\begin{figure*}[b]
  \includegraphics[height=.4\textheight]{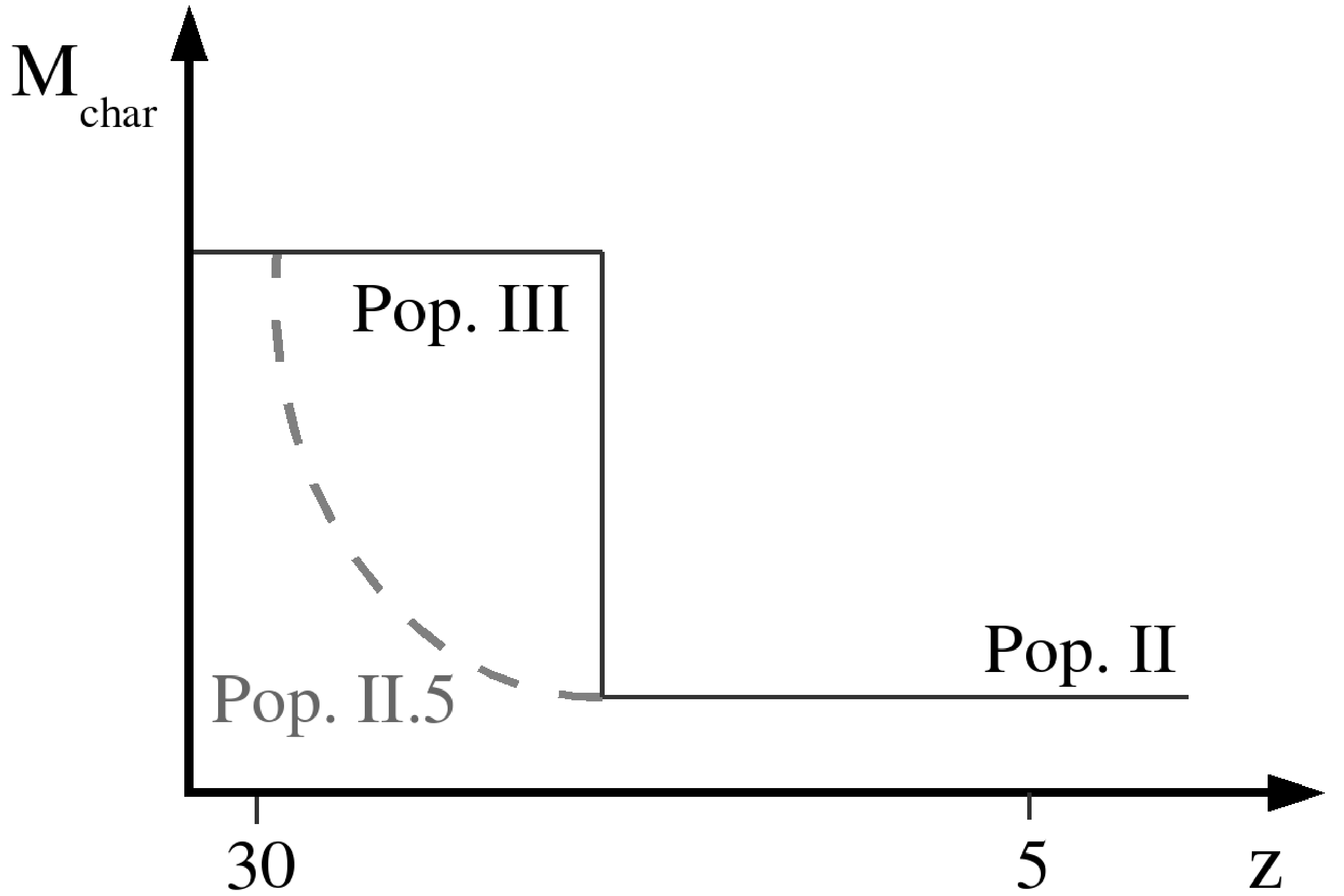}
\caption{Characteristic mass of stars as a function of redshift.
Pop~III stars, formed out of metal-free gas, and not going through
a fully ionized phase prior to the onset of collapse, have typical
masses of $M_{\ast}\sim 100 M_{\odot}$.
Pop~II stars, formed out of already metal-enriched gas, are formed at
lower redshifts, and 
have typical
masses of $M_{\ast}\sim 1 M_{\odot}$.
Pop~II.5 stars, formed out of strongly shocked, but still virtually
metal-free gas, are hypothesized to have typical masses that are
intermediate between Pop~III and Pop~II with
$M_{\ast}\sim 10 M_{\odot}$.
(Adapted from Johnson \& Bromm 2005.)}
\end{figure*}

How and when did the transition take place from the
early formation of massive stars to that of low-mass stars at later
times?  
In contrast to the formation mode of
massive stars (Population~III) at high redshifts, fragmentation is observed
to favor stars below a solar mass (Population~I and II) in the present-day
universe.  The transition between these fundamental modes is expected to be
mainly driven by the progressive enrichment of the cosmic gas with heavy
elements, which enables the gas to cool to lower temperatures.
The concept of a `critical metallicity', $Z_{\rm crit}$, has been
used to characterize the transition between
Population~III and Population~II formation modes, where $Z$ denotes the
mass fraction contributed by all heavy elements 
(Omukai 2000; Bromm et al. 2001; Schneider et al. 2002; Schneider et al. 2003;
Mackey, Bromm, \& Hernquist 2003; Yoshida, Bromm, \& Hernquist 2004).
These studies have constrained
this important parameter to only within a few orders of magnitude, $Z_{\rm
crit}\sim 10^{-6}-10^{-3} Z_{\odot}$, under the implicit assumption of
solar relative abundances of metals.
This assumption is likely to be
violated by the metal yields of the first SNe at
high-redshifts, for which strong deviations from solar abundance ratios are
predicted (e.g., Heger \& Woosley 2002; Qian \& Wasserburg 2002;
Umeda \& Nomoto 2002, 2003). 

Recently, we
have shown that the transition between the above star formation modes is
driven primarily by fine-structure line cooling of singly-ionized carbon or
neutral atomic oxygen (Bromm \& Loeb 2003).  Earlier estimates of $Z_{\rm crit}$ which
did not explicitly distinguish between different coolants are refined
by introducing
separate critical abundances for carbon and oxygen, [C/H]$_{\rm crit}$ and
[O/H]$_{\rm crit}$, respectively, where [A/H]= $\log_{10}(N_{\rm A}/N_{\rm
H})-\log_{10}(N_{\rm A}/N_{\rm H})_{\odot}$.
Since C and O are also the most important coolants
throughout most of the cool atomic ISM in present-day
galaxies, it is not implausible that these species might be
responsible for the global shift in the star formation mode.
Under the temperature and density conditions that characterize
Population~III star formation, the fine-structure lines of O\,I and
C\,II dominate over all other metal
transitions. 
Cooling due to molecules becomes important only at
lower temperatures, and cooling due to dust grains only at higher
densities (e.g., Omukai 2000,Schneider et al. 2003). 
Numerically, the critical C and O abundances are estimated to be:
[C/H]$_{\rm
crit}\simeq -3.5 \pm 0.1$ and
[O/H]$_{\rm crit}\simeq -3.1 \pm 0.2$.

\begin{figure*}[t]
  \includegraphics[height=\textwidth]{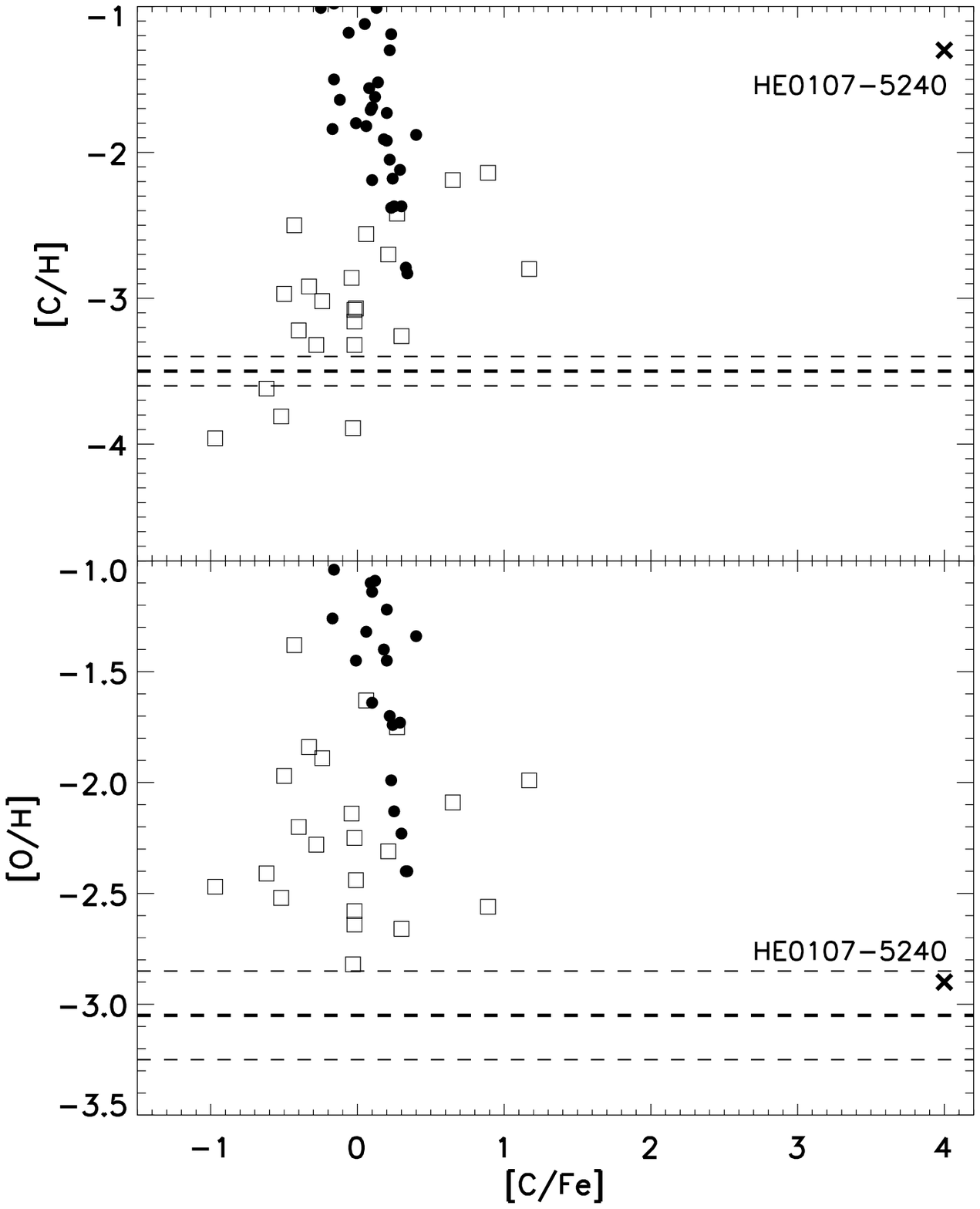}
\caption{Observed abundances in low-metallicity Galactic halo stars.
For both carbon ({\it upper panel}) and oxygen ({\it lower panel}),
filled circles correspond to samples of dwarf and subgiant stars,
and open squares to a sample of
giant stars (see \cite{BL03} for references). The dashed lines indicate
the predicted critical carbon and oxygen abundances.
Highlighted ({\it cross}) is the location
of the extremely iron-poor  giant star HE0107-5240.
(Adapted from Bromm \& Loeb 2003.)}
\end{figure*}

Even if sufficient C or O atoms are present to further cool the
gas, there will be a minimum attainable temperature
that is set by the interaction of the atoms with the thermal CMB: $T_{\rm
CMB}=2.7{\rm \,K}(1+z)$ (e.g., \cite{Lar98,CB03}).
At $z\simeq 15$, this results in a characteristic
stellar mass of $M_{\ast}\sim 20 M_{\odot}(n_f/ 10^{4}{\rm
\,cm}^{-3})^{-1/2}$, where $n_f> 10^{4}{\rm \,cm}^{-3}$ is the density at
which opacity prevents further fragmentation. 
It is possible that the transition from the high-mass to the low-mass
star formation mode was modulated by the CMB temperature and was therefore
gradual, involving intermediate-mass (`Population~II.5') stars
at intermediate redshifts (\cite{MBH03}).
This transitional population could give rise to
the faint SNe that have been proposed to explain the observed abundance
patterns in metal-poor stars (Umeda \& Nomoto 2002, 2003).
When and how uniformly the transition in the cosmic star formation
mode did take place was governed by the detailed enrichment history
of the IGM. This in turn was determined by the hydrodynamical
transport and mixing of metals from the first SN explosions (e.g.,
Mori, Ferrara, \& Madau 2002; Bromm, Yoshida, \& Hernquist 2003;
Scannapieco, Schneider, \& Ferrara 2003; Wada \& Venkatesan 2003).

Recently, the additional boost to the cooling of still metal-free
gas provided by HD has been investigated (e.g., Johnson \& Bromm 2005,
and references therein). If the primordial gas goes through a strongly
shocked, fully ionized phase prior to the onset of protostellar
collapse, cooling is possible down to the temperature of the CMB
which sets the minimum floor accessible via radiative cooling (see
Figure~3). The lower temperatures in turn could allow the fragmentation
into intermediate-mass stars, with masses of order a few tens of
$M_{\odot}$, giving rise to a possible ``Population~II.5'' (see
Figure~4).

\section{Stellar fossils}

It has long been realized that the most metal-poor stars found in
our cosmic neighborhood would encode the signature from the first stars
within their elemental abundance pattern.
For many decades, however, the observational search has failed
to discover a truly first-generation star with zero metallicity. Indeed,
there seemed to have been an observational lower limit of [Fe/H] $\sim -4$ 
(e.g., Carr 1987).
In view of the recent theoretical prediction that most Population~III stars
were very massive, with associated lifetimes of $\sim 10^6$~yr, the failure
to find any `living' Population~III star in the Galaxy is not surprising,
as they would all have died a long time ago.
Furthermore, theory has
predicted that star formation out of extremely low-metallicity gas,
with $Z\le Z_{\rm crit}\sim 10^{-3.5}Z_{\odot}$, would
be essentially equivalent to that out of truly primordial gas.
Again, this theoretical prediction was in accordance with the apparent
observed lower-metallicity cutoff.

Recently, however, this simple picture has been challenged by the
discovery of the star HE0107-5240 with a mass of $0.8 M_{\odot}$
and an {\it iron} abundance of ${\rm [Fe/H]} = -5.3$ (Christlieb et al. 2002;
Frebel et al. 2005). This finding indicates that at least
some low mass stars could have formed out of extremely low-metallicity gas.
Does the existence of this star invalidate the theory of a metallicity
threshold for enabling low-mass star formation?  
A possible explanation (Umeda \& Nomoto 2003) could lie in
the unusually high abundances of carbon and oxygen in HE0107-5240.

In Figure~5, the theoretical C and O thresholds (Bromm \& Loeb 2003)
are compared to the observed 
abundances in metal-poor dwarf  
and giant stars in
the halo of our Galaxy
(see Bromm \& Loeb 2003 for references). 
As can be seen, all data points lie above the
critical O abundance but a few cases lie below the critical C threshold.
All of these low mass stars are consistent with the model since the
corresponding O abundances lie above the predicted threshold.  The
sub-critical [C/H] abundances could have either originated in the
progenitor cloud or from the mixing of CNO-processed material (with
carbon converted into nitrogen) into the stellar atmosphere during the red
giant phase. 
Note that the extremely iron-poor
star HE0107-5240 has C and O abundances that both lie above the respective
critical levels. The formation of this low mass star ($\sim 0.8 M_{\odot}$)
is therefore consistent with the theoretical framework considered
by Bromm \& Loeb (2003).

The lessons from stellar archaeology on the nature of the first stars
are likely to increase in importance, since greatly improved, large
surveys of metal-poor Galactic halo stars are under way, or are currently
being planned.

%\begin{acknowledgments}
%\end{acknowledgments}


\begin{thebibliography}{}

%  \bibitem[van Wijngaarden (1968)]{Wijngaarden68}
%     {van Wijngaarden, L.} 1968,  
%     \textit{J.~Engng Maths} 2, 225

%\expandafter\ifx\csname natexlab\endcsname\relax\def\natexlab#1{#1}\fi
%\providecommand{\enquote}[1]{``#1''}
%\expandafter\ifx\csname url\endcsname\relax
%  \def\url#1{\texttt{#1}}\fi
%\expandafter\ifx\csname urlprefix\endcsname\relax\def\urlprefix{URL }\fi

\bibitem[]{ABN2002}Abel, T., Bryan, G. L., \& Norman, M. L. 2002, Science, 295, 93
\bibitem[]{BL2001}Barkana, R., \& Loeb, A. 2001, Physics Reports, 349, 125
\bibitem[]{BCL1999}Bromm, V., Coppi, P. S., \& Larson, R. B. 1999, ApJ, 527, L5
\bibitem[]{BCL2002}Bromm, V., Coppi, P. S., \& Larson, R. B. 2002, ApJ, 564, 23
\bibitem[]{BFCL01}Bromm, V., Ferrara, A., Coppi, P. S., \& Larson, R. B. 2001, MNRAS, 328, 969
\bibitem[]{BLar04}Bromm, V., \& Larson, R. B. 2004, ARA\&A, 42, 79
\bibitem[]{BL03}Bromm, V., \& Loeb, A. 2003, Nature, 425, 812
\bibitem[]{BL04} Bromm, V., \& Loeb, A. 2004, New Astronomy, 9, 353
\bibitem[]{BYH03}Bromm, V., Yoshida, N., \& Hernquist, L. 2003, ApJ, 596, L135
\bibitem[]{Carr87}Carr, B. J. 1987, Nature, 326, 829
\bibitem[]{Cen03}Cen, R. 2003, ApJ, 591, L5
\bibitem[]{CB03}Clarke, C. J., \& Bromm, V. 2003, MNRAS, 343, 1224
\bibitem[]{Cr02}Christlieb, N., et al. 2002, Nature, 419, 904
\bibitem[]{Fr05}Frebel, A., et al. 2005, Nature, 434, 871
\bibitem[]{FL03}Furlanetto, S. R., \& Loeb, A. 2003, ApJ, 588, 18
\bibitem[]{HW02}Heger, A., \& Woosley, S. E. 2002, ApJ, 567, 532
\bibitem[]{JB05}Johnson, J. L., \& Bromm, V. 2005, MNRAS, submitted
(astro-ph/0505304)
\bibitem[]{Kap02}Kaplinghat, M., Chu, M., Haiman, Z.,
Holder, G., Knox, L., \& Skordis, C. 2003, ApJ, 583, 24
\bibitem[]{Lar98}Larson, R. B. 1998, MNRAS, 301, 569
\bibitem[]{Lar03}Larson, R. B. 2003, Rep. Prog. Phys., 66, 1651
\bibitem[]{MBH03}Mackey, J., Bromm, V., \& Hernquist, L. 2003, ApJ, 586, 1
\bibitem[]{MFM02}Mori, M., Ferrara, A., \& Madau, P. 2002, ApJ, 571, 40
\bibitem[]{NaU2001}Nakamura, F., \& Umemura, M. 2001, ApJ, 548, 19
\bibitem[]{Om00}Omukai, K. 2000, ApJ, 534, 809
\bibitem[]{OI2002}Omukai, K., \& Inutsuka, S. 2002, MNRAS, 332, 59
\bibitem[]{ON1998}Omukai, K., \& Nishi, R. 1998, ApJ, 508, 141
\bibitem[]{OP2001}Omukai, K., \& Palla, F. 2001, ApJ, 561, L55
\bibitem[]{OP2003}Omukai, K., \& Palla, F. 2003, ApJ, 589, 677
\bibitem[]{PSS1983}Palla, F., Salpeter, E. E., \& Stahler, S. W. 1983, ApJ, 271, 632
\bibitem[]{QW02}Qian, Y.-Z., \& Wasserburg, G. J. 2002, ApJ, 567, 515
\bibitem[]{Rip2002}Ripamonti, E., Haardt, F., Ferrara, A., \& Colpi, M.
2002, MNRAS, 334, 401
\bibitem[]{SSF03}Scannapieco, E., Schneider, R., \& Ferrara, A. 2003, ApJ, 589, 35
\bibitem[]{Sch02} Schneider, R., Ferrara, A., Natarajan, 
P., \& Omukai, K.\ 2002, ApJ, 571, 30 
\bibitem[]{Sch03} Schneider, R., Ferrara, A., Salvaterra, R., 
Omukai, K., \& Bromm, V.\ 2003, Nature, 422, 869
\bibitem[]{Tan2003}Tan, J. C., \& McKee, C. F. 2004, ApJ, 603, 383
\bibitem[]{UN02}Umeda, H., \& Nomoto, K. 2002, ApJ, 565, 385
\bibitem[]{UN03}Umeda, H., \& Nomoto, K. 2003, Nature, 422, 871
\bibitem[]{WV03}Wada, K., \& Venkatesan, A. 2003, ApJ, 591, 38
\bibitem[]{WC1987}Wolfire, M. G., \& Cassinelli, J. P. 1987, ApJ, 319, 850
\bibitem[]{WyL03a}Wyithe, J. S. B., \& Loeb, A. 2003, ApJ, 588, L69
\bibitem[]{Yos2003}Yoshida, N., Abel, T., Hernquist, L., \& Sugiyama, N.
\bibitem[]{YBH2004}Yoshida, N., Bromm, V., \& Hernquist, L.
2004, ApJ, 605, 579


\end{thebibliography}
\end{document}